\documentclass[aps,twocolumn,floatfix, showpacs, superscriptaddress, nofootinbib]{revtex4}
\usepackage{amsmath}
\usepackage{graphicx,bm,hhline}
\usepackage{calrsfs}
\DeclareMathAlphabet{\pazocal}{OMS}{zplm}{m}{n}

\newcommand{\bcri}{{\bm R}_i}

\newcommand{\br}{{\bm r}}
\newcommand{\brp}{{\br}^\prime}

\newcommand{\hf}{\frac{1}{2}}

\newcommand{\nepa}{n_e^{ PA}(r)}

\newcommand{\wig}[1]{\mathrel{\hbox{\hbox to 0pt{\lower.6ex\hbox{$\sim$}\hss}\raise.4ex\hbox{$#1$}}}}

\newcommand{\intrp}{\int\,d^3r^\prime\,}

\begin{document}
\title{Electron transport calculations in warm dense matter using scattering cross sections}
\author{D. J. Burrill}
\affiliation{Los Alamos National Laboratory, P.O. Box 1663, Los Alamos, NM 87545, U.S.A.}
\author{D. Feinblum}
\affiliation{Los Alamos National Laboratory, P.O. Box 1663, Los Alamos, NM 87545, U.S.A.}
\affiliation{Department of Chemistry, University of California, Irvine, CA 92697, U.S.A.}
\author{M. R. J. Charest}
\affiliation{Los Alamos National Laboratory, P.O. Box 1663, Los Alamos, NM 87545, U.S.A.}
\author{C. E. Starrett}
\email{starrett@lanl.gov}
\affiliation{Los Alamos National Laboratory, P.O. Box 1663, Los Alamos, NM 87545, U.S.A.}

\date{\today}
\begin{abstract}
The Ziman formulation of electrical conductivity is tested in warm and hot dense matter 
using the pseudo-atom molecular dynamics method.  Several implementation options that
have been widely used in the literature are systematically tested through a comparison
to accurate but expensive Kohn-Sham density functional theory molecular dynamics (KS-DFT-MD) calculations.
The comparison is made for several elements and mixtures and for a wide range of temperatures and
densities, and reveals a preferred method that generally gives very good agreement with 
the KS-DFT-MD results, but at a fraction of the computational cost.
\end{abstract}
\maketitle

\section{Introduction}
The calculation of electron transport coefficients in warm and hot dense matter
has has been the subject of a considerable amount of recent theoretical interest 
\cite{faussurier15, reinholz15, french14, lambert11, starrett12a, hu14, faussurier14}
due in part to the importance of such quantities for inertial confinement fusion
modeling \cite{hu14, hammel10}.  A lack of direct experimental data for
electrical and thermal conductivities also drives the need for accurate and
reliable theoretical calculations.

Kohn-Sham density functional theory molecular dynamics\footnote{Hereafter referred
to as quantum molecular dynamics (QMD).} has emerged as a powerful and accurate tool
for simulating warm dense matter.  By coupling it with the Kubo-Greenwood
formulation \cite{greenwood58, kubo57, hanson11, desjarlais02} it has also been 
used to generate conductivity data.  Unfortunately,
the method is extremely computationally expensive, and for this reason is
typically limited to low temperatures (a few tens of eV).  

A much more computationally efficient alternative to QMD are average atom models.
These come in many varieties \cite{sterne07, blenski, piron3, perrot87, liberman, scaalp, rozsnyai}
and are also density functional theory based models.  The common theme to these models is that
one tries to find the properties of one atom in the plasma.  Average atom models have also
been used to generate electronic transport coefficients.  Broadly speaking,
there are to two main approaches: on the one hand are Kubo-Greenwood
based calculations \cite{johnson09, johnson, johnson2, starrett12a} that
are qualitatively similar to the QMD calculations;
and on the other hand are Ziman-type methods that estimate conductivity on the basis
of electron-ion scattering cross sections \cite{sterne07, perrot87, faussurier15, pain10, dharma06, rozsnyai08}.

Previously \cite{starrett12a}, the average atom Kubo-Greenwood approach was systematically compared to QMD data for
a number of elements and mixtures for a range of temperatures and densities.  The comparison
revealed reasonable agreement in some cases, but some serious disagreement in others, both qualitative 
and quantitative.  In this paper we adopt a similar systematic approach but using the Ziman-type
method.  While much has been published using this approach, a systematic evaluation of its accuracy relative
to QMD simulations has not, to the best of our knowledge, been carried out.  Moreover, to generate
the inputs (i.e. scattering potential, ionization fraction and ion-ion structure factor) for the
Ziman-type approach we use the recently developed pseudo-atom molecular dynamics (PAMD) method \cite{starrett15}.
PAMD has been shown to be of comparable accuracy to DFT-MD methods for equation of state, ionic diffusion
coefficients and ion-ion structure factors \cite{starrett13, starrett15}, though it has limitations \cite{starrett14b}. 
The advantage of of PAMD over DFT-MD methods is that it is much less computationally expensive.
As a result of the accuracy of PAMD, the comparison of electron conductivities
to QMD should be a fair test of the conductivity model (i.e. the Ziman-type formula), and not be seriously
limited by inaccuracies in the underlying model.

In section \ref{sec_exp} we review the Ziman-Evans formula for electronic conductivity, and then rewrite
it in terms of an electronic relaxation time.  This allows the connection and relationship of the Ziman-Evans formula 
to the recently developed expression of Johnson \cite{johnson09} to be clarified.  Moreover, we are able
to generalize Johnson's expression to explicitly include the ion-ion structure factor.  
We then give the extensions of these formulas for mixtures, and in the case of the Johnson formula, its 
extension to thermal and optical conductivities.  In section \ref{sec_pamd} we summarize the pseudo-atom
molecular dynamics method and discuss the options for coupling it to the conductivity formulas. 
Then in section \ref{sec_res} the systematic comparison to the QMD data is carried out.
We also test
the accuracy of Thomas-Fermi PAMD compared to Kohn-Sham PAMD for warm dense aluminum.
Finally in section \ref{sec_conc} we draw our conclusions.

\section{Expressions for electron transport coefficients \label{sec_exp}}
In this section we review the commonly used Ziman-Evans formula for resistivity.  We rewrite it in terms
of an electronic relaxation time in line with Ziman's original derivation, and give the relationship
between the direct approach to the conductivity (as originally proposed by Ziman \cite{ziman61})
and the inverse approach, where the inverse conductivity (i.e. the resistivity) is directly 
calculated (as proposed by Evans \cite{evans73}).  Both approaches have recently been used
in the literature (eg. by Johnson \cite{johnson09} and Faussurier {\it et al} 
\cite{faussurier15}) without discussion of their relative merits or relationship.
Next we show how these formulations are easily extended to mixtures and finally,
how the direct approach is easily extended to optical conductivity and thermal
conductivity.  We use Hartree atomic units in which $\hbar = m_e = e = 1$.

\subsection{Ziman-Evans formulation}
The Ziman-Evans formula for calculating the resistivity $R$, which is the inverse of the conductivity,
is \cite{faber72, evans73, perrot87}
\begin{equation}
\label{eq:r}
R = -\frac{n_I^0}{3 \pi (n_e^*)^2 }
\int\limits_0^{\infty} d\epsilon  \frac{\partial f}{\partial\epsilon} 
\int\limits_{0}^{2 p} q^3 \frac{\partial \sigma(\epsilon,\theta)}{\partial \theta} S_{II}(q) dq
\end{equation}
where $\epsilon = p^2 / 2 $, 
$n_e^*$ is the number density of scattering electrons, $n_I^0$ is the density of ions,
$S_{II}(k)$ is the ion-ion static structure factor, 
\begin{equation}
q^2 = 2p^2[1-\cos\theta]
\end{equation}
is the magnitude of momentum transfer squared, and
\begin{equation}
f = \frac{1}{\exp\left( \beta (\epsilon-\mu_e^*) \right) + 1 }
\end{equation}
is the Fermi-Dirac occupation factor.  Here $\beta = 1 /T $ is the inverse temperature
and $\mu_e^*$ is the electron chemical potential.
The differential cross section is given by
\begin{equation}
\label{eq:tmatrix}
\frac{\partial \sigma(\epsilon,\theta)}{\partial \theta}  = \frac{1}{p^2}
\left| \sum\limits_{l=0}^{\infty} (2l+1) \sin \eta_l e^{\imath \eta_l} P_l(\cos\theta) \right|^2
\end{equation}
where $\eta_l$ are phase shifts and $P_l(\cos\theta)$ are 
Legendre polynomials.  Equation~\eqref{eq:tmatrix} is
referred to as the transition- or $t$-matrix approach to the cross section.
The phase shifts $\eta_l$ in equation (\ref{eq:tmatrix}) are
determined by solving the Schroedinger equation for a scattering potential $V^{scatt}(r)$.

In the Born approximation to the differential cross section we have
\begin{equation}
\frac{\partial \sigma(\epsilon,\theta)}{\partial \theta}  \approx \frac{1}{4\pi^2} \tilde{V}^{scatt}(q)^2
\end{equation}
Using this approximation and  integrating Eq.~\eqref{eq:r} by parts gives~\cite{ichimaru85}
\begin{equation}
R = \frac{n_I^0}{12 \pi^3 (n_e^*)^2 }
\int\limits_0^{\infty} dk k^3 f(k/2) \tilde{V}^{scatt}(k)^2 S_{II}(k)
\end{equation}
where $\tilde{V}^{scatt}(k)$ is the Fourier transform of $V^{scatt}(r)$.

\subsection{Relaxation time formulation}
Following Ziman \cite{ziman61} and Johnson \cite{johnson09} we define a relaxation time $\tau_p$
\begin{equation}
\frac{1}{\tau_p} = \pi n_I^0\,\frac{v}{p^4}\, 
\int_0^{2p} dq\, q^3 \frac{d\sigma}{d\theta}(p,\theta) S_{II}(q)
\end{equation}
where $p=m_e v$.  This agrees with the relaxation time used in \cite{johnson09} except that we have included
a structure factor, whereas Johnson's formula has $S_{II}(q) = 1$.
We can now write down the conductivity $\sigma_{DC}$ using the Boltzmann approach of Ziman \cite{ziman61} 
and the resistivity $R$ using \cite{evans73} 
\begin{equation}
\sigma_{DC} = -\frac{1}{3 \pi^2 } \int_0^\infty d\epsilon \frac{df}{d\epsilon} v^3 \tau_p
\label{direct}
\end{equation}
\begin{equation}
R     = -\frac{1}{3 \pi^2 (n_e^*)^2} \int_0^\infty d\epsilon \frac{df}{d\epsilon} v^3 \frac{1}{\tau_p}
\label{inverse}
\end{equation}
As $T \to 0$,  $df/d\epsilon \to - \delta(\epsilon - \epsilon_F)$, where $\epsilon_F$ is the Fermi 
energy.  In this limit the Fermi velocity $v_F = (3 \pi^2 n_e^*)^{1/3}$
and both these expressions give the same result
\begin{equation}
\sigma_{DC} = \frac{1}{R}  = n_e^*\, \tau_{v_F}
\label{drude}
\end{equation}
i.e. the Drude form.  For $T > 0$ however, the results will differ.  We refer to equation (\ref{direct})
as the direct approach to the conductivity, and equation (\ref{inverse}) as the inverse approach.  It is easy
to show that the inverse approach and the Ziman-Evans formula (equation (\ref{eq:r})) are the same.

In recovering the Drude expression (equation (\ref{drude})) from both formulations we have assumed that the chemical potential and 
electron density are related through
\begin{equation}
n_e^* = c_{TF} I_{1/2}[\beta \mu_e^*] = \frac{1}{\pi^2}\int_0^\infty d\epsilon\, v\, f(\epsilon)
\label{nestar}
\end{equation}
where $c_{TF}$ is a constant \cite{starrett13} and $I_{1/2}$ is the Fermi integral of order $1/2$ \cite{starrett13}.  
These equations (\ref{direct}), (\ref{inverse}) and (\ref{nestar}) implicitly assume a free electron
density of states.


\subsection{Extension to mixtures}
To extend the Ziman-Evans formulation to mixtures we follow ref. \cite{dwp1}.  The 
extension is simply achieved by the replacement:
\begin{equation}
\frac{\partial \sigma(\epsilon,\theta)}{\partial \theta} S_{II}(q) \to
\sum_{i}^{N_s} \sum_{j}^{N_s} (x_i x_j)^{\hf} S_{ij}(q)
\pazocal{F}_i(\epsilon,\theta)
\pazocal{F}_j^\star(\epsilon,\theta)
\end{equation}
where $N_s$ is the number species and $\pazocal{F}_i$ is the scattering amplitude for species $i$
\begin{equation}
\pazocal{F}_i(\epsilon,\theta)
 = \frac{1}{p}
\sum\limits_{l=0}^{\infty} (2l+1) \sin \eta_l^i e^{\imath \eta_l^i} P_l(\cos\theta)
\end{equation}
or, in the Born approximation
\begin{equation}
\pazocal{F}_i(\epsilon,\theta)
 = \frac{V_i^{scatt}(q)}{2\pi}
\end{equation}
$x_i$ is the number fraction of species $i$ and
$S_{ij}$ are the partial structure factors.

\subsection{Extensions to the direct approach}
Ziman \cite{ziman60} (Ziman's equation 9.9.7) gives the expression for conductivity in terms of the integral expressions
\begin{equation}
K_n = -\frac{1}{3 \pi^2 } \int_0^\infty d\epsilon \frac{df}{d\epsilon} (\epsilon-\mu_e^*)^n v^3 \tau_p
\end{equation}
so that $\sigma_{DC} = K_0$.  He also gives the expression for the thermal
conductivity $\kappa$ in terms of these integrals
\begin{equation}
\kappa = \frac{1}{T}\left( K_2 - \frac{(K_1)^2}{K_0}  \right)
\label{kappa}
\end{equation}

For completeness we also give the extension of equation (\ref{direct}) to non-zero frequencies,
but do not present any calculations based on it.  Following Johnson \cite{johnson09},
\begin{equation}
\sigma(\omega) = -\frac{1}{3 \pi^2 } \int_0^\infty d\epsilon \frac{df}{d\epsilon} v^3 \frac{\tau_p}{(\omega \tau_p)^2 + 1}
\label{loww}
\end{equation}
This gives the free-free contribution to the optical conductivity and uses a low-frequency approximation.  
This formula satisfies the conductivity sum rule
\begin{eqnarray}
\int_0^\infty d\omega \sigma(\omega) & = & \frac{1}{2 \pi} \int_0^\infty d\epsilon\, v\, f(\epsilon) \nonumber\\
                                     & = & \frac{\pi}{2} n_e^*
\end{eqnarray}

To our knowledge, such extensions have not been developed for the inverse
approach, though an expression for the inverse thermal conductivity in the Born approximation has
been given in ref. \cite{ichimaru85}.

\subsection{Physical interpretation of the quantities in the Ziman-like equations}
\begin{figure}
\begin{center}
\includegraphics[scale=0.35]{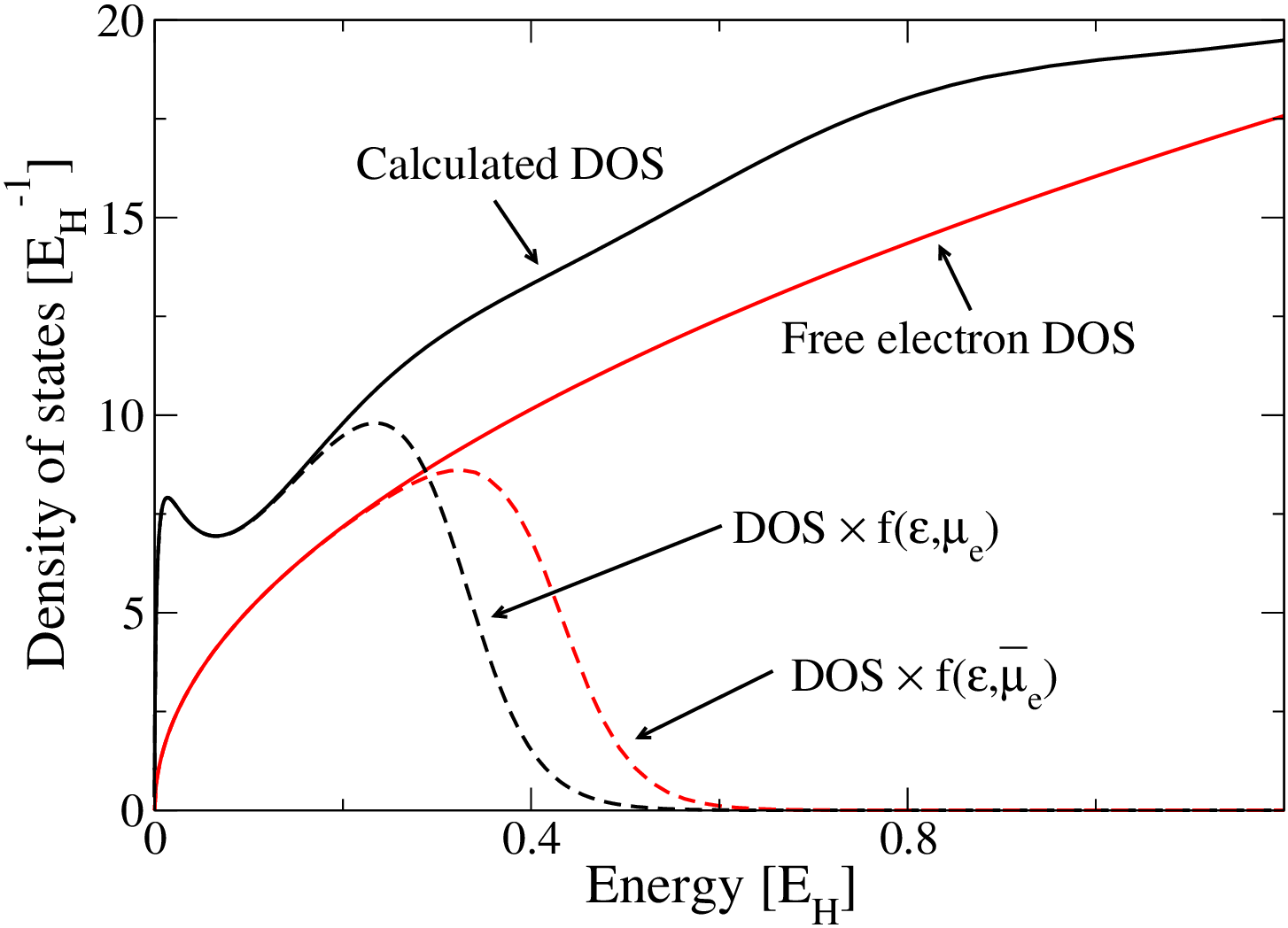}
\end{center}
\caption{(Color online) An example of a calculated density of states (DOS) compared to
a free electron DOS for aluminum at solid density and 1 eV.  
The dashed lines are the ``filled'' states, i.e. the DOS
multiplied by the Fermi-Dirac occupation factor.  The area under either dashed
curve is the number of ionized electrons per atom, both curves integrate to 3,
as expected for aluminum under these conditions.  
The conductivity formulas (equations (\ref{direct}) and (\ref{inverse}))
assume a free electron density of states.
}
\label{fig_dos}
\end{figure}
The above Ziman-like equations (\ref{direct}) and (\ref{inverse}) for resistivity require input of the potential 
from which the electron scatters $V^{scatt}(r)$, 
the density of scatterers $n_e^*$, the ion-ion structure factor $S_{II}(k)$
and the electronic chemical potential $\mu_e^*$.
In Ziman's original derivation \cite{ziman61}, $V^{scatt}(r)$ is a weak pseudopotential such that the Born
approximation is valid and the total potential $V^{tot}(\br)$ is given by the superposition of $V^{scatt}(r)$ at each
atomic site, i.e.
\begin{equation}
V^{tot}(\br) = \sum_i V^{scatt}(|\br - \bcri|)
\label{zsuper}
\end{equation}
and $n_e^*$ is the density of valence electrons.  Evans \cite{evans73} assumed the same form (\ref{zsuper})
but $V^{scatt}(r)$ was allowed to be a strong scattering potential, not a pseudopotential.  He further required that
the scattering potential be of muffin-tin type, such that $V^{scatt}(r)=0$ for $r> R_{MT}$, where $R_{MT}$ is
some muffin-tin radius.  Formally, it is required that no two muffin-tin potentials overlap.  The pseudopotential
in Ziman's formulation is supposed to preserve the scattering properties of the full potential in Evans'
approach.

For $n_e^*$ there are at least two reasonable choices.  If the number of ionized electrons per ion is $\bar{Z}$,
then the density of these valence electron is $\bar{n}_e^0 = n_I^0 \bar{Z}$.  It seems reasonable to identify
$n_e^* = \bar{n}_e^0$.  However, the electron density $n_e^*$ is also related to the chemical potential $\mu_e^*$.
If we take the limit $T \to 0$ of the conductivity formulas (equations (\ref{direct}) and (\ref{inverse})), we should
recover the Drude form, equation (\ref{drude}).  Because of the implicit free-electron density of states (DOS) in
equations (\ref{direct}) and (\ref{inverse}), to recover the Drude form $\mu_e^*$ must be related to $n_e^*$
by equation (\ref{nestar}).  If $n_e^* = \bar{n}_e^0$ then we can use equation (\ref{nestar}) to find a corresponding
chemical potential, $\bar{\mu}_e$.  However, in general the plasmas we consider do not have a free electron
DOS.  An example of the differences between a free electron DOS and a calculated DOS is given in figure \ref{fig_dos}.
The important point is that with the calculated DOS a different chemical potential, $\mu_e$ say, is needed to recover
the correct number of valence electrons $\bar{Z}$.  If we use (the more realistic) $\mu_e$ in the conductivity formulas 
then because of their implicit free electron DOS, a different electron density, $n_e^0$ say, results from 
equation (\ref{nestar}).  Thus we have two reasonable choices: either use $\bar{n}_e^0$ and $\bar{\mu}_e$, or
$n_e^0$ and $\mu_e$.  In what follows we test both choices.  Finally we note that this ambiguity is brought
about by the implicit free electron DOS in the conductivity formulas and would be resolved if a calculated DOS
could be used in equations (\ref{direct}) and (\ref{inverse}).  However, we do not attempt this here.

\begin{figure}
\begin{center}
\includegraphics[scale=0.35]{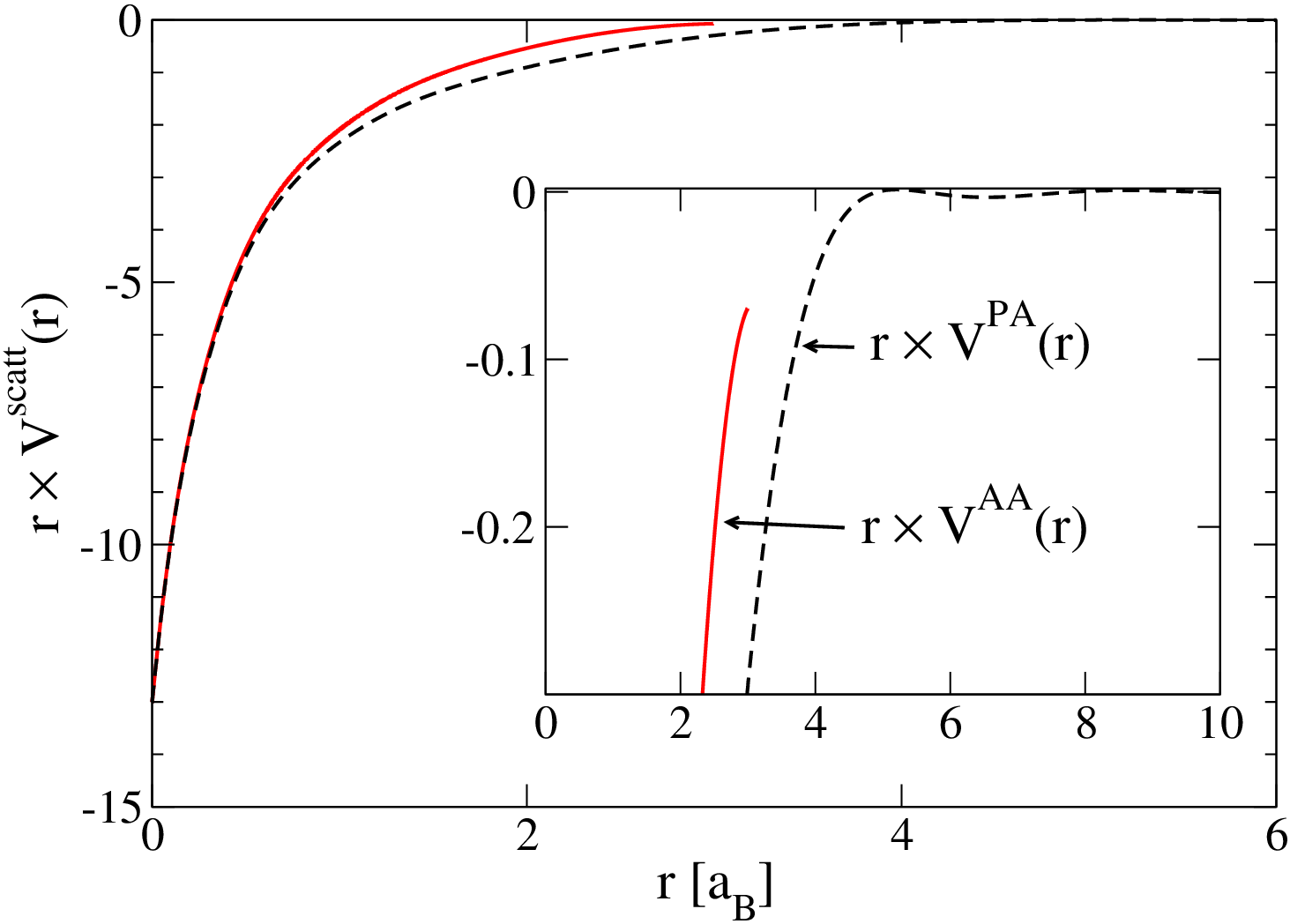}
\end{center}
\caption{(Color online) An example of the potentials $V^{AA}$ and $V^{PA}$
for aluminum at solid density and 1 eV.  The muffin-tin like $V^{AA}$ 
extends as far as the ion-sphere radius $R_{WS}$ ($\approx 3$ $a_B$ for aluminum at
this density).  The potential $V^{PA}$ is not muffin-tin like and can have long range 
structure due to Friedel oscillations in the electron density. 
The inset shows the same potentials, but focused on the regions where
the potentials are close to 0.
}
\label{fig_potentials}
\end{figure}

\section{Underlying physical model \label{sec_pamd}}
The Ziman-like formulas require inputs of $V^{scatt}(r)$, $n_e^*$, $\mu_e^*$ and $S_{II}(k)$
from an underlying physical model.  Numerous models have been used in the literature, and
each can be coupled to the Ziman-like formulas in different ways, depending on what
one chooses for the inputs.  Here,
for our underlying model we choose the recently developed Pseudo-atom molecular
dynamics (PAMD) \cite{starrett15}.  This model self consistently calculates the
electronic and ionic structure, and it agrees well with generally accepted DFT-MD
methods on equation of state, ionic transport and the structure factor itself \cite{starrett15, starrett14}.
Moreover, it agrees well with a recent X-ray scattering experiment that is sensitive
to the ion-ion structure factor \cite{starrett15b}.  Thus we have confidence in the inputs 
that it generates for the Ziman-like formulas.  In the next section we summarize PAMD with
the aim of describing the various options for how it can be coupled to the Ziman-like
formulas.

\begin{table*}
\begin{center}
\begin{tabular}{ l l }
\hline
\hline
 Input choice   & Description \\
\hline
 $V^{scatt}(r)$   & \\
 \hspace{0.5cm} $V^{AA}$   & Equation (\ref{vaa}).  The muffin-tin like average atom potential. \\
 \hspace{0.5cm} $V^{PA}$   & Equation (\ref{vpa}).  The potential due to one pseudo-atom. \\
 \hspace{0.5cm} $V^{scr}$  & Equation (\ref{vscr}).  The screened pseudopotential. \\
\hline
 $\mu_e^*$, $n_e^*$   & \\
 \hspace{0.5cm} $\mu_e$, $n_e^0$               & The ``physical'' quantities calculated in the average atom model. The correct number of valence \\
 \hspace{0.5cm}                                & electrons is returned if $\mu_e$ is used with the calculated, non-free electron density of states.\\
 \hspace{0.5cm} $\bar{\mu}_e$, $\bar{n}_e^0$   & The quantities corresponding the density of valence electrons if a free-electron density of states is used. \\
\hline
\hline
\end{tabular}
\caption{\label{tab:1} Input choices for the Ziman-like formulas that are tested here.
}
\end{center}
\end{table*}

\subsection{Summary of pseudo-atom molecular dynamics}
In PAMD the electron density of the plasma $n_e(\br)$ is approximated
by a superposition of identical pseudo-atom electron densities $n_e^{PA}(r)$
placed at each nuclear site $\bcri$
\begin{equation}
n_e(\br) = \sum_i n_e^{PA}(|\br - \bcri|)
\label{super}
\end{equation}
The total scattering potential is given by
\begin{equation}
V^{tot}(\br) = \sum_i V^{PA}_{el}(|\br - \bcri|) + V^{xc}_{ee}[n_e(\br)]
\label{supervpa}
\end{equation}
where
\begin{equation}
V^{PA}_{el}(r) = \frac{-Z}{r} + \intrp \frac{n_e^{PA}(r^\prime)}{\left| \brp - \br \right|} 
\label{vpael}
\end{equation}
$Z$ is the nuclear charge of the pseudo-atom, and $V_{ee}^{xc}$ is the exchange and correlation
potential.  The ionic positions $\bcri$
are generated in PAMD using classical molecular dynamics (MD), with a pseudo-atom
pair interaction potential $V(r)$ being calculated directly from $n_e^{PA}(r)$.
Alternatively, if one requires only knowledge of the ion-ion structure factor,
$V(r)$ can be used in the so-called Ornstein-Zernike (OZ) equations.  Solution
of the OZ equations is much faster than using MD  and
is very accurate provided a good approximation for the bridge function is 
used\footnote{For the bridge function have used VMHNC for single species plasmas \cite{rosenfeld86, faussurier04} and HNC 
for mixtures \cite{starrett14b}, since VMHNC is not available.  See \cite{starrett15b} for further details.}.
We have used the OZ solution for all results presented here.

The calculation of $V(r)$ has been described in detail in refs. \cite{starrett13,starrett14}, but it is
relevant to summarize it here.  One starts by defining an electron density $n_e^{scr}(r)$,
that is the component of $n_e^{PA}(r)$ due solely to valence (or screening) electrons.
In Fourier space, $V(k)$ is given by
\begin{equation}
V(k) = \frac{4\pi}{k^2}\bar{Z}^2 - n_e^{scr}(k) \frac{C_{ie}(k)}{\beta}
\end{equation}
where $\bar{Z} = \int\, d^3r\, n_e^{scr}(r)$ and $-C_{ie}(k)/\beta$ is often called
a pseudopotential \cite{perrot87}.  This pseudopotential is generated by requiring that in the
linear response regime $-C_{ie}(k)/\beta$ is the potential that returns the
density $n_e^{scr}$, i.e.
\begin{equation}
n_e^{scr}(k)  = - \frac{C_{ie}(k)}{\beta} \chi_e(k)
\end{equation}
where $\chi_e$ is the response function of the electrons (see equation (17) of \cite{starrett14}).
This type of pseudopotential has been used, along with Born approximation, to calculate resistivities
with the Ziman-Evans formula \cite{perrot87, ichimaru85}.  While this pseudopotential is weak by construction
(due to the linear response relation), it is not designed to preserve the scattering
properties of the full scattering potential ($V^{PA}(r)$ in our case, see equation (\ref{vpa})).  In the usual pseudopotential
formulation (the one assumed by Ziman \cite{ziman61}) one tries to preserve the phase shifts of the full scattering potential, 
and modify the potential only near the nuclei.
This is not guaranteed for $-C_{ie}(k)/\beta$ with potentially serious consequences for the resulting scattering 
cross sections.  Nevertheless, we have implemented this calculation and
show results using it in section \ref{sec_res}.  It is important to note that it is the screened pseudopotential $V^{scr}(r)$
that is used in the Ziman-like formulas, where
\begin{equation}
V^{scr}(k)  = - \frac{C_{ie}(k)}{\epsilon(k)\,\beta} 
\label{vscr}
\end{equation}
where $\epsilon(k)$ is the dielectric function \cite{perrot87}.


The process for calculating $\nepa$ involves solving an average atom (AA) model. 
In this model, we use density functional theory to find the electron density $n_e^{full}(r)$ 
\footnote{We retain the notation used in \cite{starrett14} with the exception that $V_{Ne}^{eff}$
is renamed $V^{AA}$.} in a spherically symmetric 
model system, in which a nucleus of charge $Z$ is placed at the origin and the remaining ions
in the system are modeled as a continuous distribution represented by a Heaviside step function $\Theta(r-R_{WS})$
at the Wigner-Seitz radius $R_{WS}$ (this is called the ion-sphere model).   A sphere of radius $R_{WS}$ surrounding
the central nucleus is required to the charge neutral.  The total average atom potential is given by
\begin{eqnarray}
  V^{AA}(r) & = & -\frac{Z}{r} + \int d\brp \frac{(n_e^{full}(r^\prime) - n_e^0 \Theta(r^\prime - R_{WS}) )}{\mid\br-\brp\mid} \nonumber\\
                     && + V_{ee}^{xc}[n_e^{full}(r)] -  V_{ee}^{xc}[n_e^0] 
\label{vaa}
\end{eqnarray}
where $n_e^0$ is the electron density in the field free region.  The average atom chemical potential $\mu_e$
is determined when solving the model, and is related to $n_e^0$ by equation (\ref{nestar}).  It is a good estimation of the
``physical'' chemical potential, as it takes into account a non-free electron density of states (see figure \ref{fig_dos}).   

$V^{AA}(r)$ can be
thought of as a muffin-tin potential, in the sense that Evans \cite{evans73} meant.  It violates the condition
that no-two muffin-tins overlap, however this may not be too serious a problem since such overlaps are routinely tolerated 
when using muffin-tin potentials for X-ray spectroscopy calculations \cite{natoli03}.  However it is fair to point
out that the average atom potential is not generated in the same way as muffin-tin potentials are usually generated \cite{natoli03}.

The screened potential (\ref{vscr}) and the average atom potential (\ref{vaa}) have been widely used with the Ziman-Evans
formula to calculate conductivities \cite{perrot87, pain10, hansen06, rinker88, rozsnyai08}.  A third type of potential, that of a single
pseudo-atom, has also been used \cite{faussurier15, perrot87}.  It is important to note that while all these previous calculations
have used potentials falling broadly into these three categories, differences in implementation details are important and have
significant consequences on conductivity calculations.  In PAMD we can define a pseudo-atom potential $V^{PA}(r)$ as
\begin{equation}
V^{PA}(r) = V^{PA}_{el}(r) + V_{xc}^{PA}(r)
\label{vpa}
\end{equation}
where the electrostatic part $V_{el}^{PA}$ was defined in equation (\ref{vpael}).  The exchange and correlation part $V_{xc}^{PA}$
cannot be exactly separated from $V_{ee}^{xc}[n_e(\br)]$ (equation (\ref{supervpa})), but a reasonable approximation can be written down in
analogy with the definition of the pseudo-atom electron density.  $\nepa$ is calculated from the 
average atom electron density $n_e^{full}(r)$ by removing the electron density found in a system with 
the same surrounding ions, but no central nucleus, $n_e^{ext}(r)$ (see \cite{starrett14} for further details):
\begin{equation}
  n_e^{PA}(r) = n_e^{full}(r) - n_e^{ext}(r) \label{pa}
\end{equation}
Therefore we define
\begin{equation}
  V_{xc}^{PA}(r) \equiv V_{ee}^{xc}[n_e^{full}(r)] - V_{ee}^{xc}[n_e^{ext}(r)]
\end{equation}
An example of $V^{AA}(r)$ and $V^{PA}(r)$ is given in figure \ref{fig_potentials}.  Note that
$V^{AA}(r)$ does not go exactly to zero at $R_{WS}$ (the muffin-tin radius), which is consistent
with typical muffin-tin calculations \cite{peyrusse08}.

In what follows we will numerically test all three potentials and both types of chemical potential.  We will
also demonstrate the effect of the ion-ion structure factor.  All of these potentials, and the structure factor
are self-consistently calculated in PAMD.  Due to agreement with widely accepted DFT-MD simulations on equation
of state, ionic self-diffusion and static ionic structure, as well as with available experimental measurements,
we have a high degree of confidence in the inputs to the Ziman-like formulas generated by PAMD.  Through a 
systematic comparison of these various inputs with QMD conductivity data we aim to evaluate the best choice
of inputs, if such a choice does indeed exist.  We restrict ourselves to comparisons where we have a high degree
of confidence in the PAMD model.  Roughly speaking this means temperatures $> 1$ eV and densities roughly
equal to or greater than solid density.  It may be possible to go to lower densities or temperatures but
that would require further testing and possible improvements to the PAMD model.
As an aid to the reader we summarize the various input options that are to be tested in table \ref{tab:1}

\begin{figure}
\begin{center}
\includegraphics[scale=0.4]{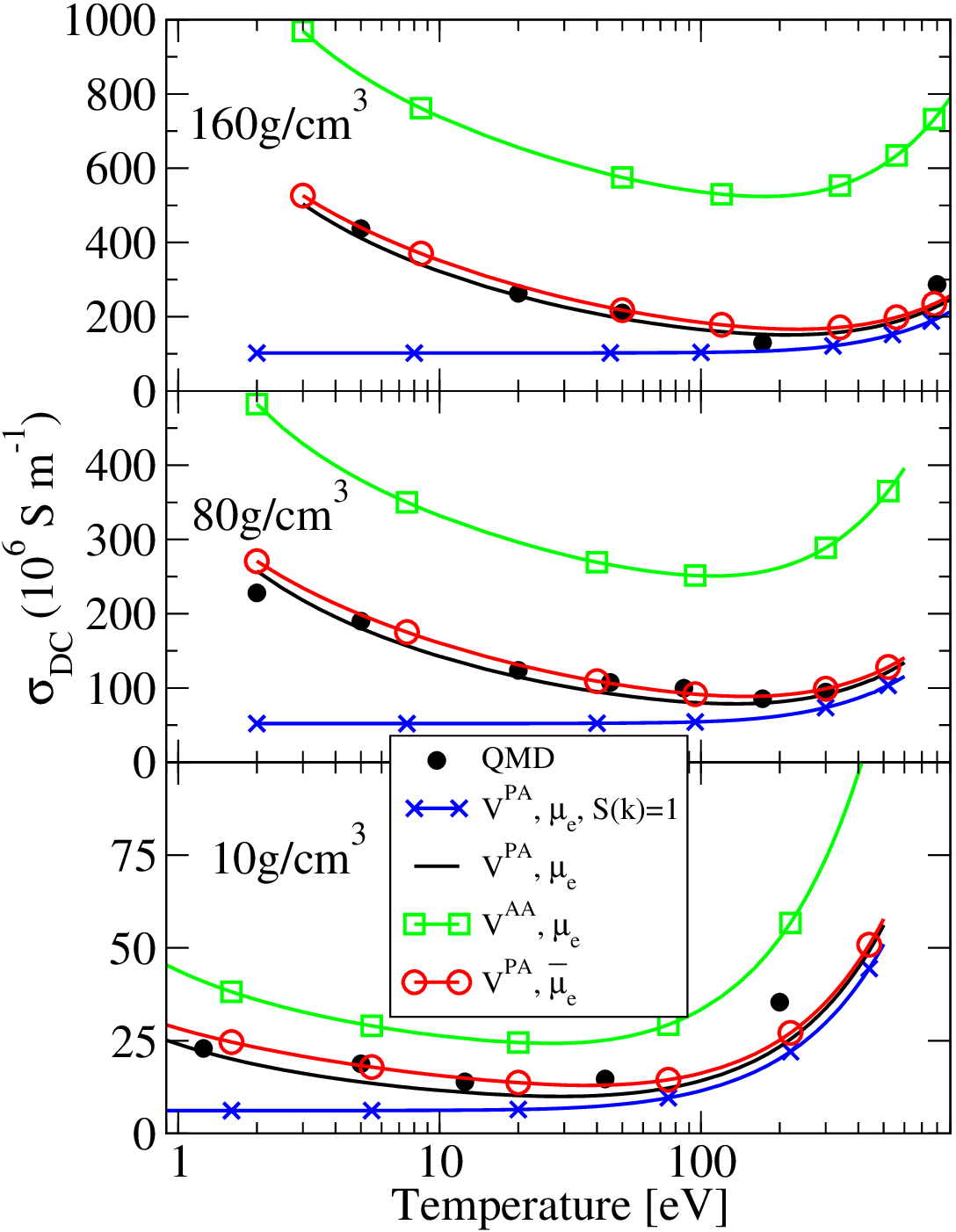}
\end{center}
\caption{(Color online) Electrical conductivity of dense hydrogen.  All curves use PAMD-KS with a $t$-matrix calculation
 of the conductivity.  The QMD data is from \cite{lambert11}.  Clearly using the average atom (AA) potential grossly
overestimates the conductivity relative to QMD, while setting $S(k)=1$ demonstrates that the increase in conductivity
on lowering the temperature is due to changes in ionic structure.}
\label{fig_h1}
\end{figure}

\begin{figure}
\begin{center}
\includegraphics[scale=0.4]{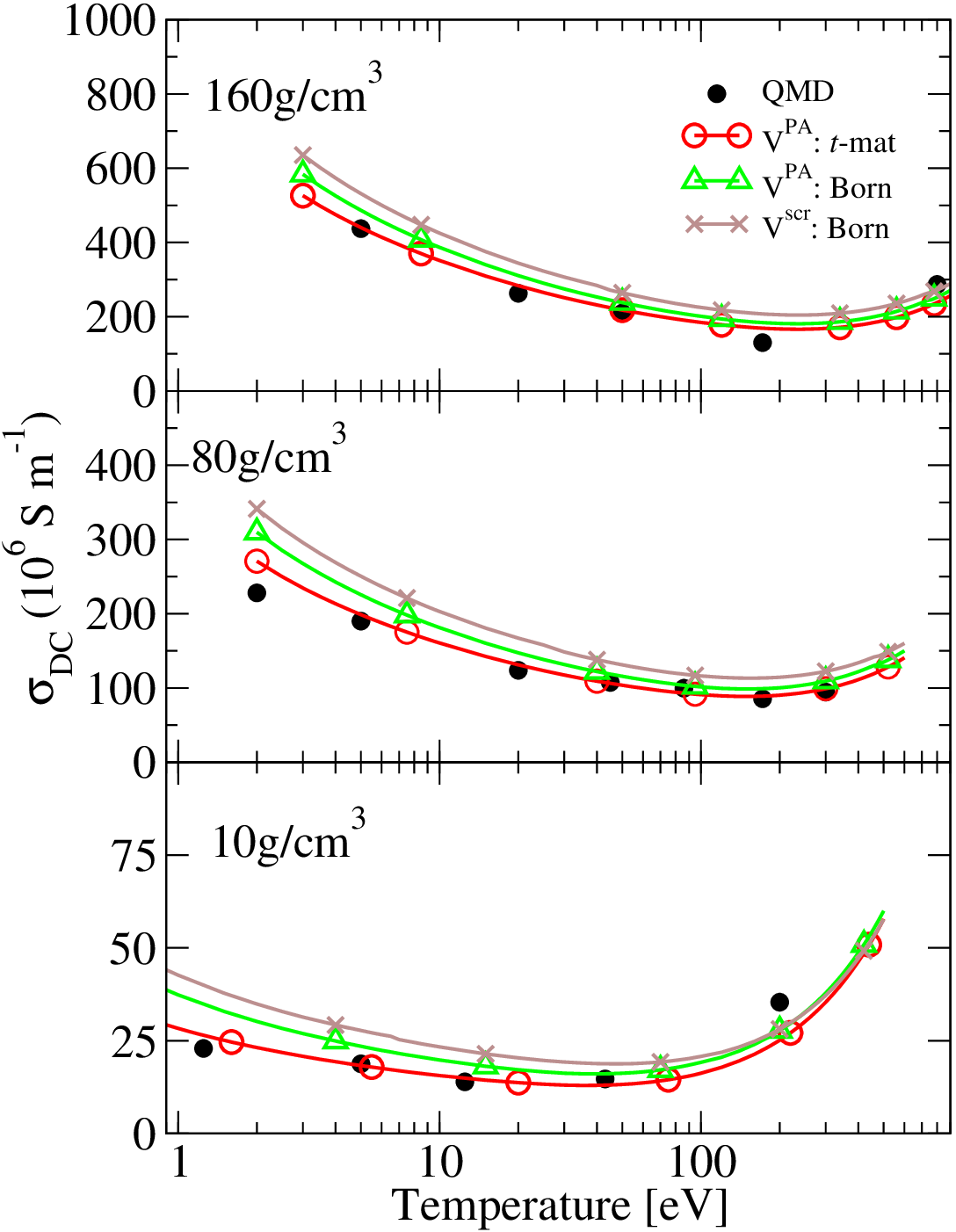}
\end{center}
\caption{(Color online) Electrical conductivity of dense hydrogen.  All curves use PAMD-KS with chemical potential $\bar{\mu}_e$.  
The QMD data is from \cite{lambert11}.  The Born calculations for both potentials are reasonably close to the QMD results. 
}
\label{fig_h2}
\end{figure}

\begin{figure}
\begin{center}
\includegraphics[scale=0.35]{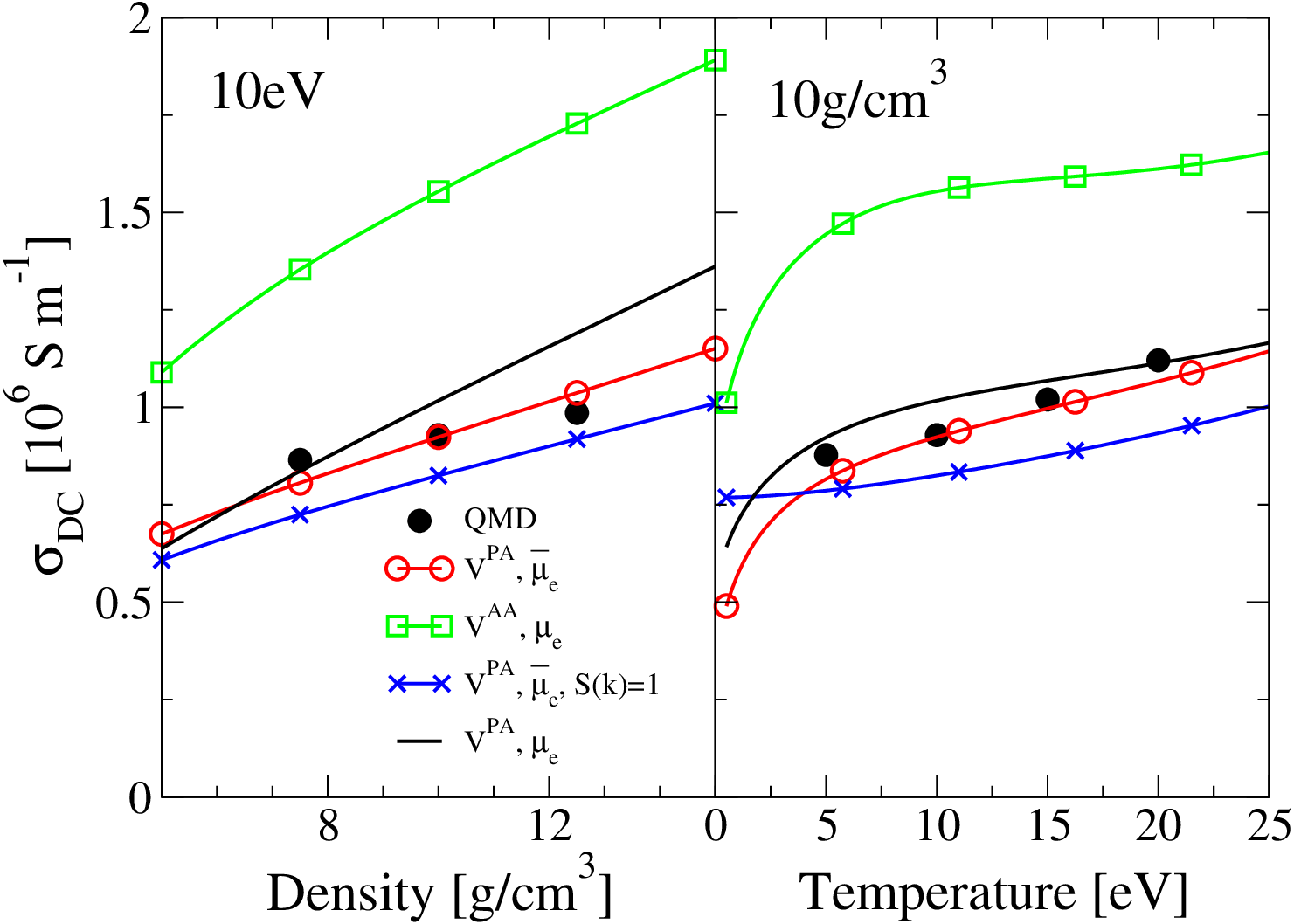}
\end{center}
\caption{(Color online) Electrical conductivity of dense beryllium.  
All curves use PAMD-KS with a $t$-matrix calculation of the conductivity.  
The QMD data is from \cite{starrett12a}.  
As for hydrogen, using the average atom potential overestimates the conductivity.
}
\label{fig_be1}
\end{figure}

\begin{figure}
\begin{center}
\includegraphics[scale=0.35]{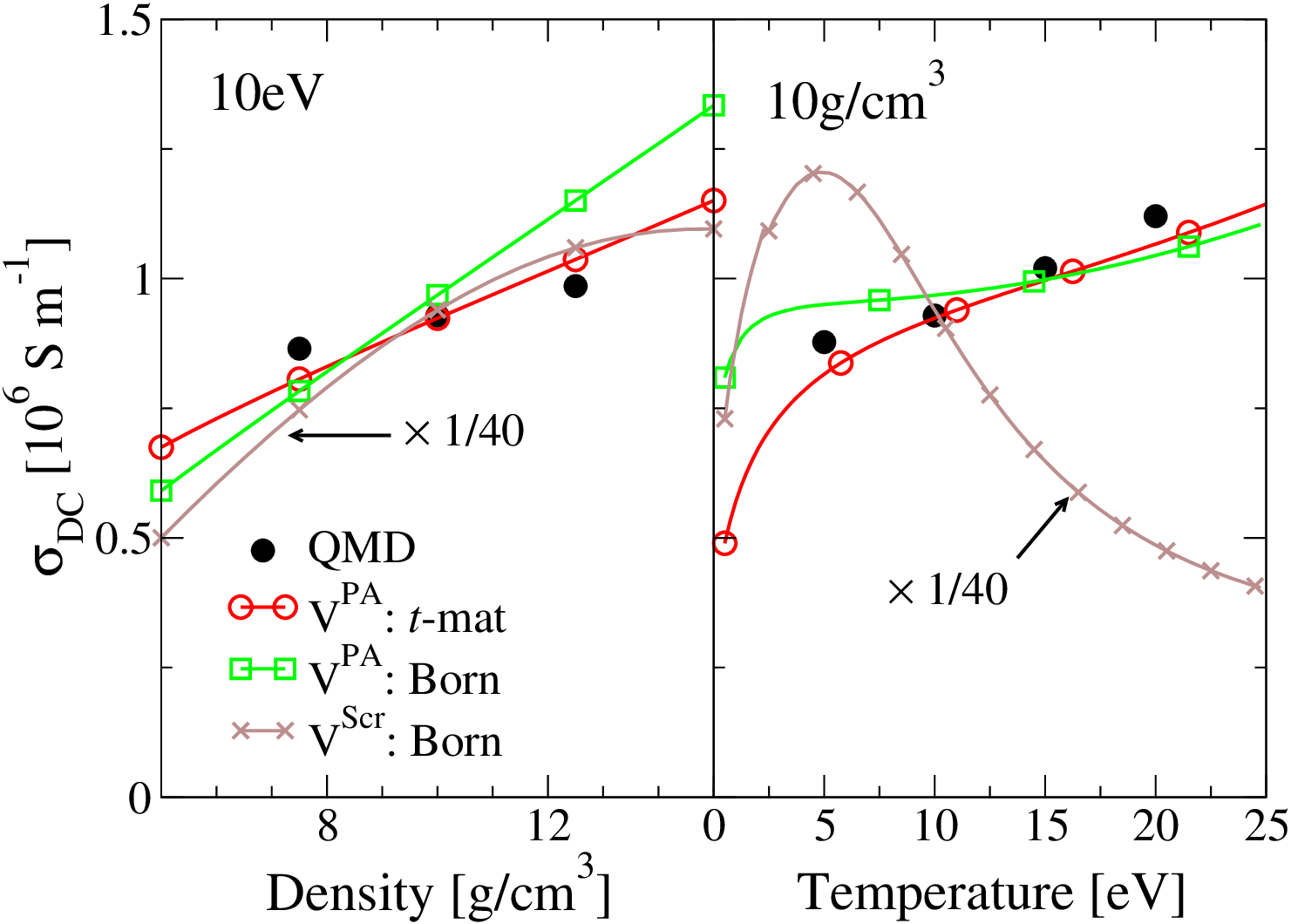}
\end{center}
\caption{(Color online) Electrical conductivity of dense beryllium.  
All curves use PAMD-KS with chemical potential $\bar{\mu}_{e}$.  
The QMD data is from \cite{starrett12a}.  
While the Born approximation for the $V^{PA}$ potential gives reasonable
results, that for $V^{scr}$ strongly disagrees in magnitude with the QMD
data (it was scaled by a factor of 1/40, as indicated in the plot).  
}
\label{fig_be2}
\end{figure}

\section{comparison of results \label{sec_res}}
\begin{figure}
\begin{center}
\includegraphics[scale=0.4]{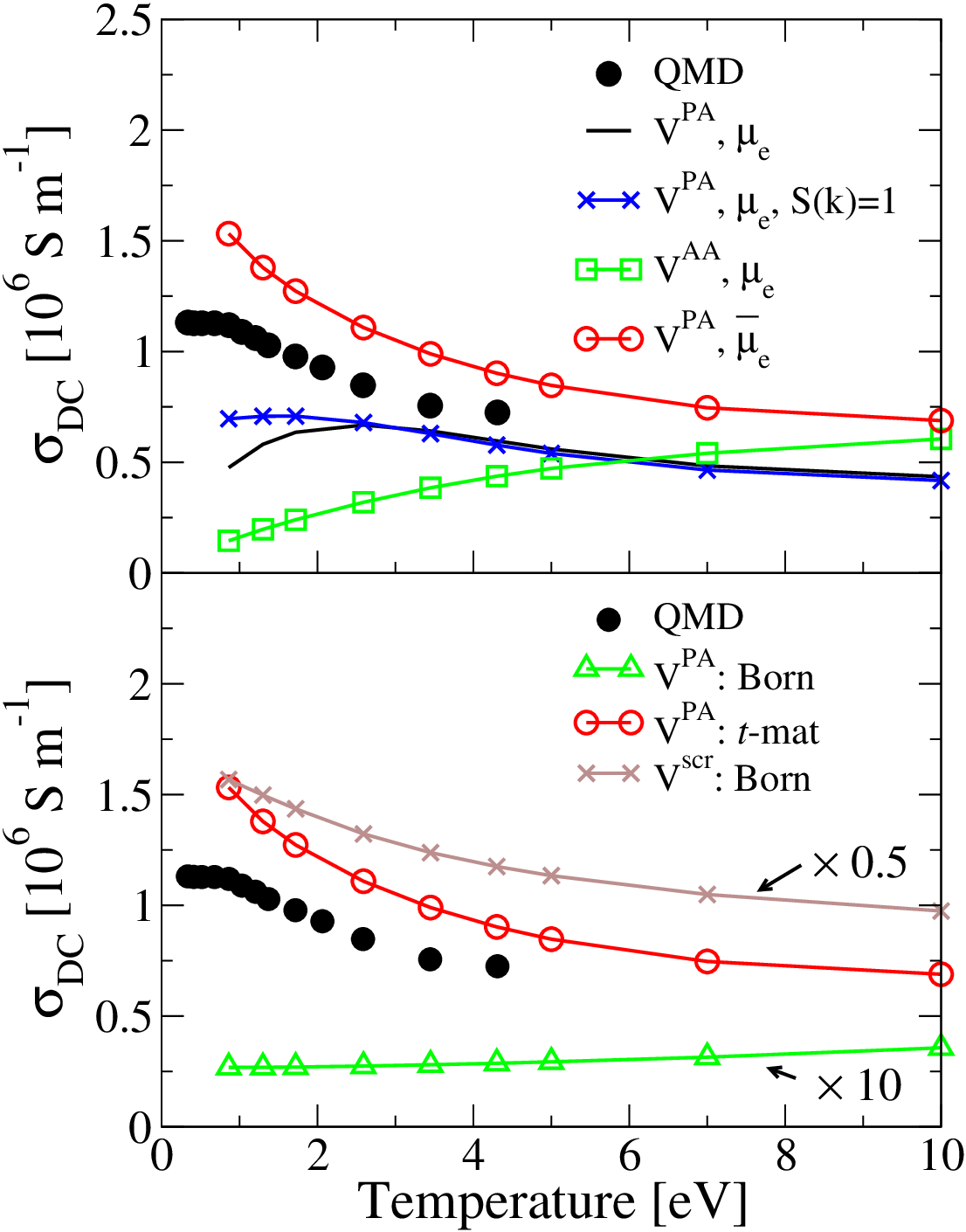}
\end{center}
\caption{(Color online) Electrical conductivity of molybdenum at 10 g/cm$^3$.  
All curves use PAMD-KS.  In the top panel all calculations use the $t$-matrix
method, in the bottom all calculations use the chemical potential $\bar{\mu}_{e}$.
The QMD data is from \cite{french14}.  
In the bottom panel the $V^{scr}$ Born results have been scaled by the factor indicated in
the plot.
}
\label{fig_mo}
\end{figure}

\begin{figure}
\begin{center}
\includegraphics[scale=0.4]{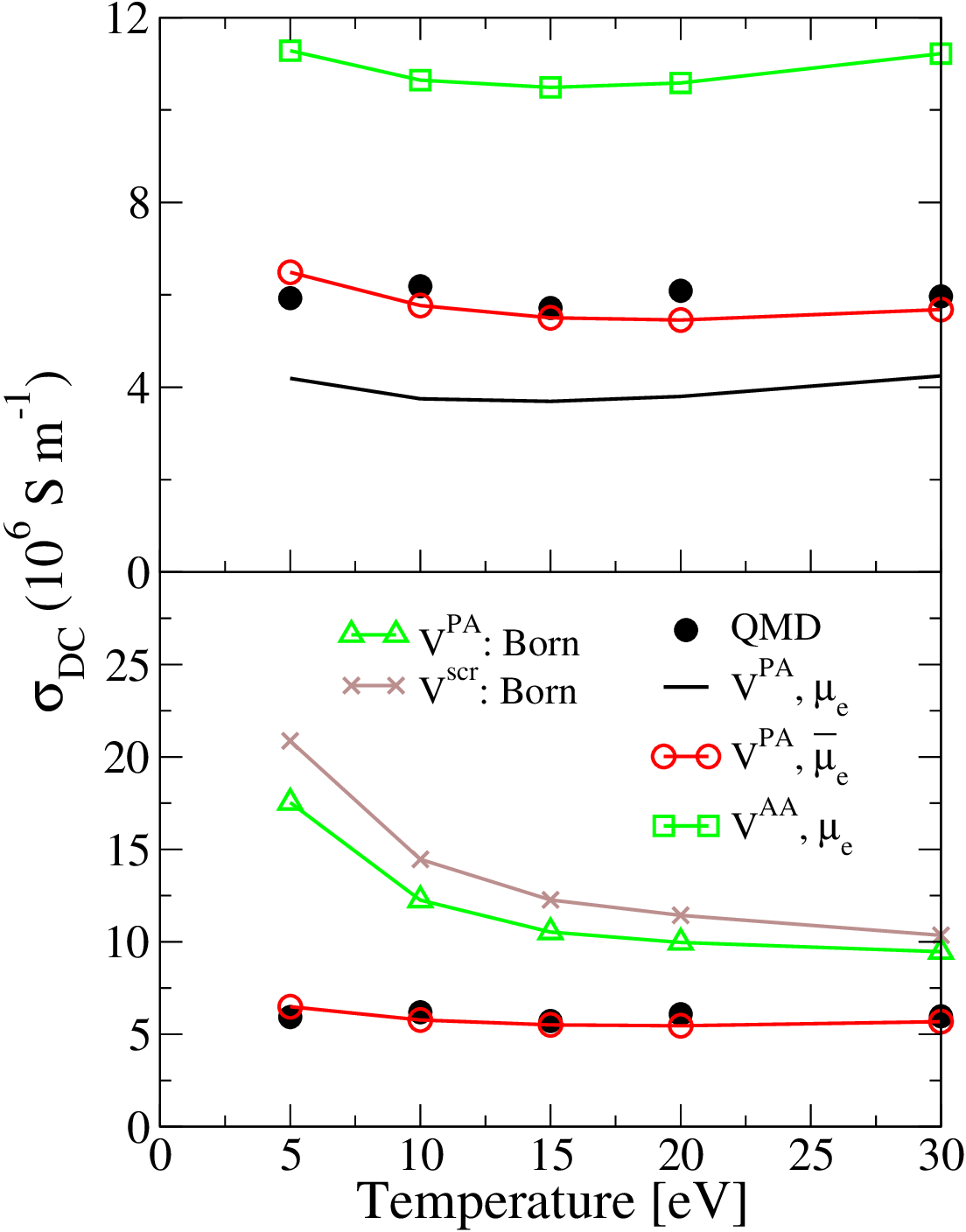}
\end{center}
\caption{(Color online) Electrical conductivity of a deuterium/tritium mixture (1:1) at 10 g/cm$^3$.  
All curves use PAMD-KS.  In the top panel all calculations use the $t$-matrix
method, in the bottom all calculations use the chemical potential $\bar{\mu}_{e}$.
The QMD data is from \cite{starrett12a}.  
}
\label{fig_dt}
\end{figure}

\begin{figure}
\begin{center}
\includegraphics[scale=0.4]{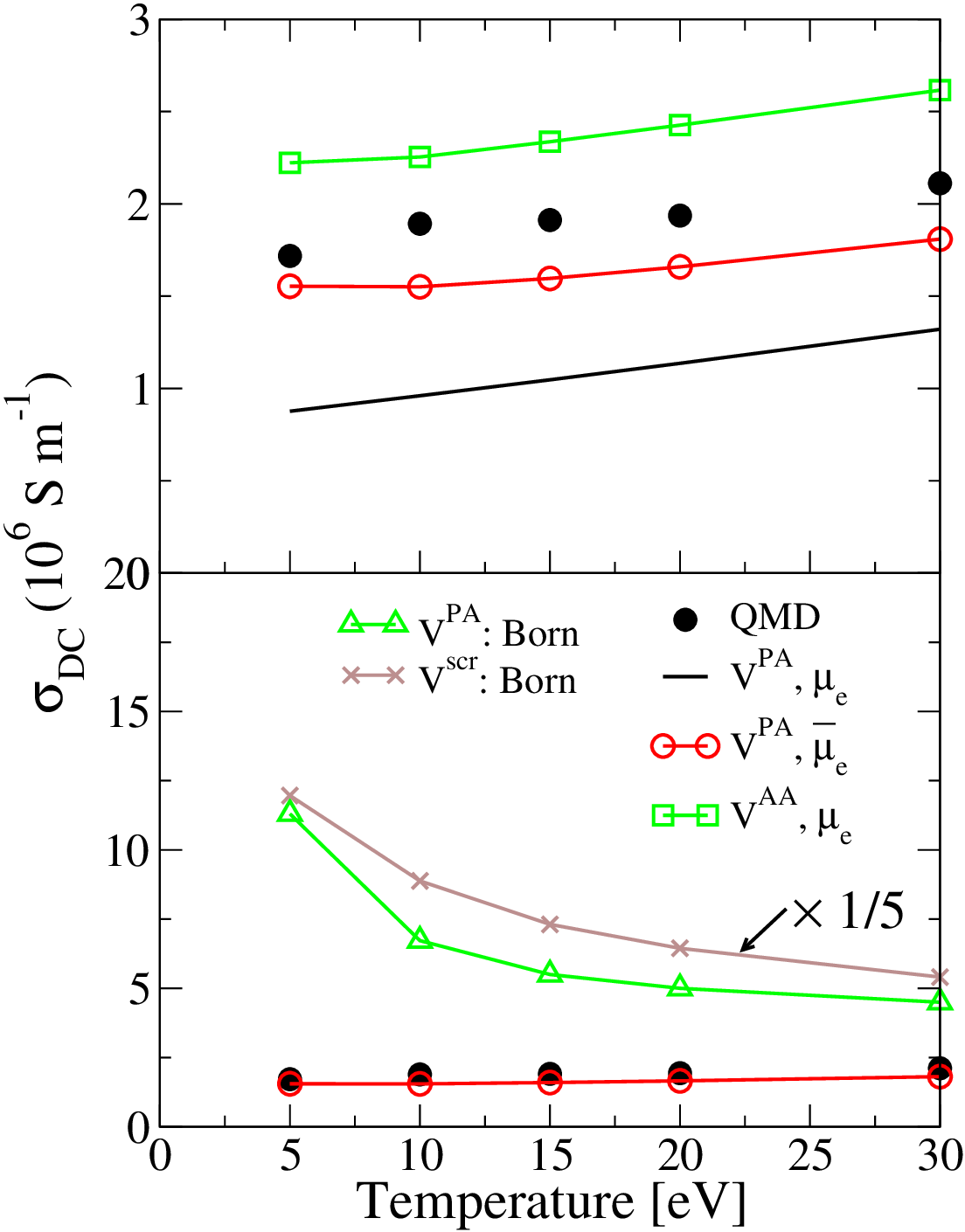}
\end{center}
\caption{(Color online) Electrical conductivity of a carbon/hydrogen mixture (7:9) at 10 g/cm$^3$.  
All curves use PAMD-KS.  In the top panel all calculations use the $t$-matrix
method, in the bottom all calculations use the chemical potential $\bar{\mu}_{e}$.
The QMD data is from \cite{starrett12a}.  
In the bottom panel the Born results have been scaled by the factors indicated in
the plot.
}
\label{fig_ch}
\end{figure}

\begin{figure}
\begin{center}
\includegraphics[scale=0.4]{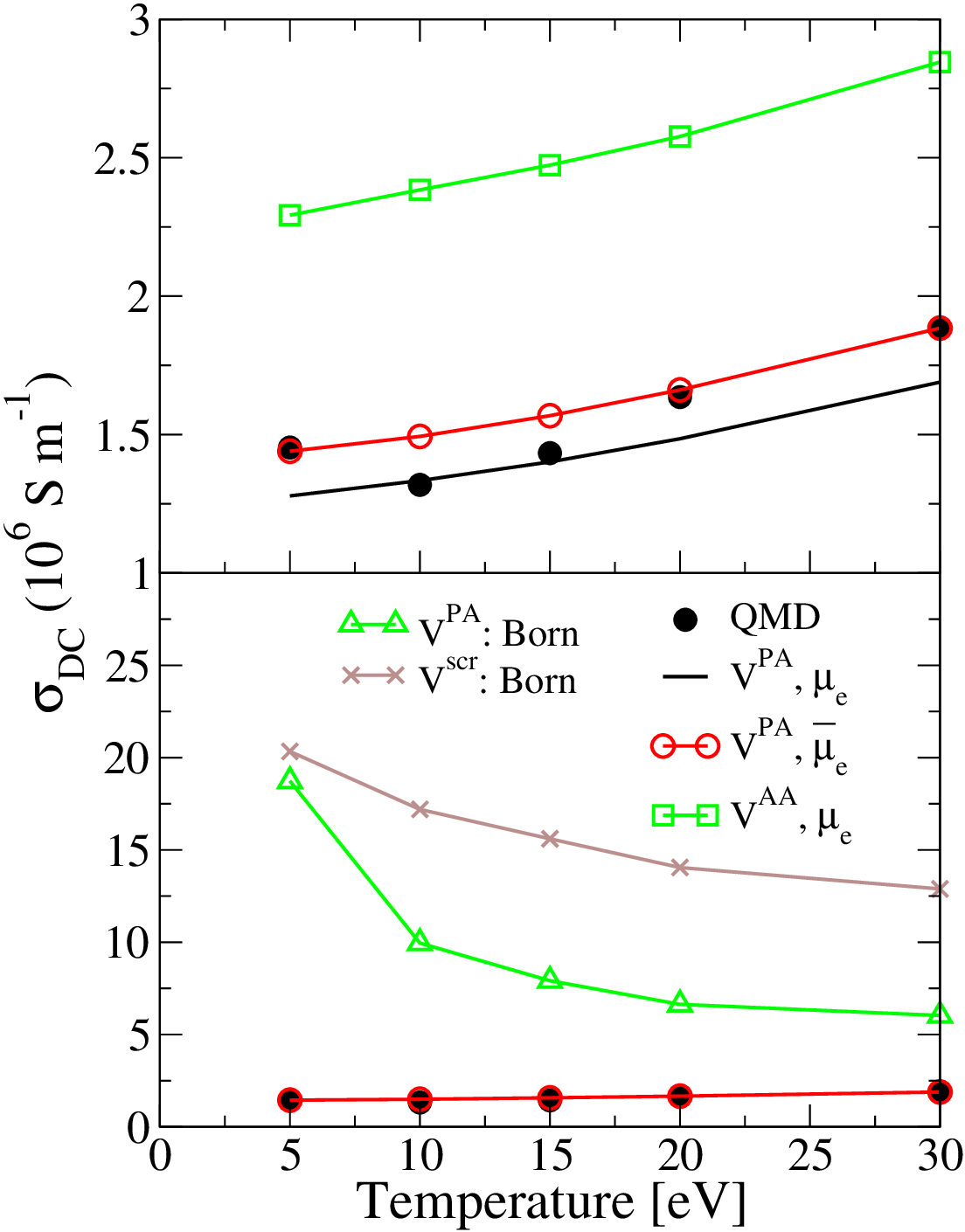}
\end{center}
\caption{(Color online) Electrical conductivity of a beryllium/deuterium/tritium mixture (1:1:1) at 10 g/cm$^3$.  
All curves use PAMD-KS.  In the top panel all calculations use the $t$-matrix
method, in the bottom all calculations use the chemical potential $\bar{\mu}_{e}$.
The QMD data is from \cite{starrett12a}.  
}
\label{fig_bedt}
\end{figure}

PAMD is a density functional theory based model.  In it we
can use either Kohn-Sham (KS) or Thomas-Fermi (TF) treatments of the electronic
kinetic energy term.  The KS treatment is more physically accurate and in figures
\ref{fig_h1} to \ref{fig_bedt} we restrict ourselves to it.  We
must also choose an exchange and correlation potential, for which we have used the finite
temperature local density approximation of \cite{ksdt}.  Also, in figures \ref{fig_h1}
to \ref{fig_bedt} we use the inverse approach to conductivity only (equation (\ref{inverse})).
We address Thomas-Fermi results and the direct approach in sections \ref{sec_tf}
and \ref{sec_direct}.
We compare to QMD calculations and assume that these are close to the correct result.
QMD is thought to be an accurate method for determining the electrical conductivity
in warm dense matter, but is very expensive computationally.  There can be issues with
convergence of these calculations \cite{lambert11, recoules3} due to the extreme
computational expense, but in the absence of experiments, they are widely accepted as being
the most accurate data available.

In figure \ref{fig_h1} we show the electrical conductivity
of dense hydrogen.  We have used the $t$-matrix method for all curves.
Clearly, using the AA potential results in significantly too large
a conductivity compared to QMD, though qualitatively the results with $V^{AA}$ are
reasonable.  Using the pseudo-atom potential $V^{PA}$ we find very good agreement with
the QMD.  For $V^{PA}$ we show results using both chemical potential options, and
find only small differences.  However, in the physical regimes shown in the plot,
the two estimates of $n_e^*$ are close.  Setting the structure factor to be equal to
1, we see that the rise in conductivity at low T is no longer reproduced.  In \cite{starrett12a}
a similar effect was seen, though not explained.  Those calculations \cite{starrett12a}
used an average atom potential and no explicit structure factor was included, thus 
implicitly using $S(k)=1$.  From figure \ref{fig_h1} we can now understand the result
of \cite{starrett12a}:  The lack of the rise in conductivity at low T was due to 
the implicit $S(k)=1$, while the overestimation was because the average atom potential
was used.

In figure \ref{fig_h2} we show the same QMD hydrogen data, but now test the Born
approximation.  Using $V^{PA}$ with the Born approximation worsens agreement with
the QMD data, though it is still reasonable.  Using the screened pseudopotential 
$V^{scr}$ with the Born approximation worsens the agreement still further.

We now turn to warm dense beryllium.  This is a more complicated atom than hydrogen
and one may expect a comparison to QMD data to be a more robust test of the method.
In figure \ref{fig_be1} we show $t$-matrix calculations.
Again, we see that using $V^{AA}$ overestimates the QMD, while using $V^{PA}$ with
either chemical potential choice gives good agreement.  We find that for the conditions
of the plot, setting $S(k)=1$ only has a small quantitative effect, though qualitatively
a decrease in the conductivity at low T seen in the full calculation disappears.
In figure \ref{fig_be2} we test the Born approximations.
While using the Born approximation with $V^{PA}$ results in reasonable agreement with 
the QMD, using it with $V^{scr}$ leads to vast differences.  To fit the curve on the same
plot we had to scale it by a factor of 1/40.  This result strongly suggests that,
even where the Born approximation is reasonable, using the screened pseudopotential
can lead to seriously inaccurate results.

Figure \ref{fig_mo} shows results for molybdenum.  This is a much more complicated atom
than hydrogen or beryllium and is a transition metal.  This comparison should 
be an even more stringent test of the method.  In the figure see that both
$V^{PA}$ and $V^{AA}$ give reasonable agreement in magnitude with the QMD.  Using
$V^{AA}$ gives poorer qualitative agreement at the lower temperatures, but we must
be cautious, since it is likely that the underlying model (PAMD) will become
less accurate at the lowest temperatures seen, where molybdenum approaches its
solid phase.  The two choices of chemical potential make a relatively large difference
here, with the two curves bracketing the QMD data.  The effect of $S(k)$ is only
important at the lowest temperatures.  In the bottom panel of the figure
we test the Born approximation.  For both potentials $V^{scr}$ and $V^{PA}$ the
results have been scaled as indicated, and both do a poor job of estimating
the conductivity.  Clearly the Born approximation has broken down here, as the
large difference between the $t$-matrix and Born results for $V^{PA}$ indicate.
The fact that using $V^{scr}$ comes closer to the data than for beryllium is
probably due to a cancellation of errors: the Born approximation underestimates $\sigma_{DC}$, while
$V^{scr}$ leads to an overestimation.

We now turn to mixtures: deuterium/tritium (1:1), figure \ref{fig_dt};
carbon/hydrogen (7:9), figure \ref{fig_ch}; beryllium/deuterium/tritium (1:1:1),
figure \ref{fig_bedt}.  The results are broadly in line with the previous
result seen in figures \ref{fig_h1} to \ref{fig_mo}.  Using $V^{PA}$ with 
either chemical potential gives good agreement with the QMD data, with the 
choice $n_e^* = \bar{n}_e^0$ leading to somewhat better agreement in general.
Using $V^{AA}$ leads to an overestimation of the conductivity, though we find
better agreement where the heavier element is present (the carbon/hydrogen mixture).
In all cases the Born approximation with $V^{PA}$ overestimates the corresponding
$t$-matrix calculation, and this overestimation becomes worse for the lower temperatures.
Using $V^{scr}$ with the Born approximation makes the agreement with QMD worse still,
and for the carbon/hydrogen case it is in particularly bad agreement.

\subsection{Discussion of the comparison to QMD}
The comparisons in figures \ref{fig_h1} through \ref{fig_bedt} indicate
that using $V^{PA}$ with either chemical potential option leads the
good agreement with QMD data.  Using $n_e^* = \bar{n}_e^0$ is perhaps
the slightly better option, with significantly better agreement being found 
in some cases.  Using $V^{scr}$ with the Born approximation seems
to be an unreliable method, sometimes leading to very inaccurate results, even
where the Born approximation is reasonable.  As discussed earlier, the
screened pseudopotential is not designed for this purpose, and is
not guaranteed to preserve the scattering properties of the full
potential ($V^{PA}$).  Since $V^{scr}$ is generated from the screening electron
density $n_e^{scr}(r)$, one possibility is that removing the oscillations
in $n_e^{scr}(r)$ due to core valence orthogonality, before generating 
$V^{scr}$, could improve the results.  We have tested this by using the 
Troullier-Martins method \cite{troullier91}, and found no significant improvement (not shown). Moreover,
the resulting conductivity becomes strongly dependent on the Troullier-Martins core radius, 
a highly undesirable quality.  In summary, we do not recommend using the $V^{scr}$
potential.

The average atom potential leads to surprisingly poor agreement in magnitude for many
cases.  This is surprising because it has been used frequently in the past 
\cite{pain10, sterne07, rozsnyai08}, and because it is consistent
with Evans' \cite{evans73} original formulation.  The failing  may be due
to the muffin-tin approximation.  In making this approximation Evans's
formula is limited to non-overlapping scattering potentials.  By forcing
the potential extend only as far as the muffin-tin radius (the Wigner-Seitz radius
for $V^{AA}$) one immediately prevents recovery of Ziman's formula, which
is correct in the Born approximation limit.  In that limit, if the scattering
potential is long ranged (i.e. longer ranged than $R_{WS}$) then the muffin-tin
approximation will significantly change it (i.e. shorten its range).   Only if the scattering potential
is shorter ranged than $R_{WS}$ does Evans' approach recover Ziman's in the Born
limit.  Using $V^{PA}$ is a compromise between these two approaches that appears
to capture the best parts of both.  The potential is not modified by the muffin-tin
approximation, so the Born limit can be recovered, while the use of the $t$-matrix
method as proposed by Evans, allows strong scatterers to be present.  Qualitatively,
it is also consistent with Ziman's original idea of having isolated scattering
centers \cite{ziman61}.

\begin{figure}
\begin{center}
\includegraphics[scale=0.4]{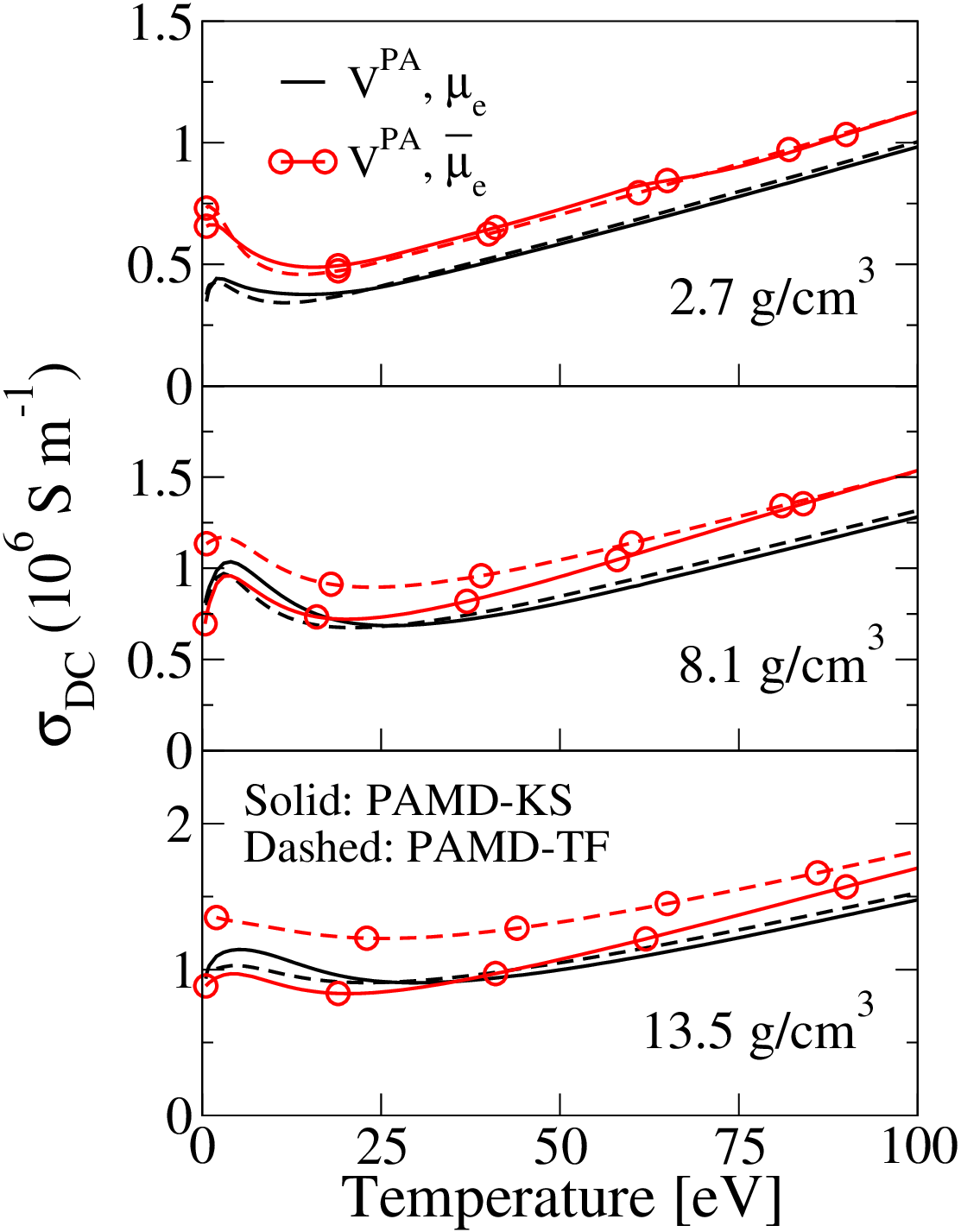}
\end{center}
\caption{(Color online) Electrical conductivity of aluminum, comparing Thomas-Fermi
(TF) to Kohn-Sham (KS) based PAMD for three different densities.  The PAMD-TF results
are generally quite close to PAMD-KS, and are very computationally efficient to calculate.
All calculations use $V^{PA}$ and the $t$-matrix approach.
}
\label{fig_tf}
\end{figure}

\begin{figure}
\begin{center}
\includegraphics[scale=0.4]{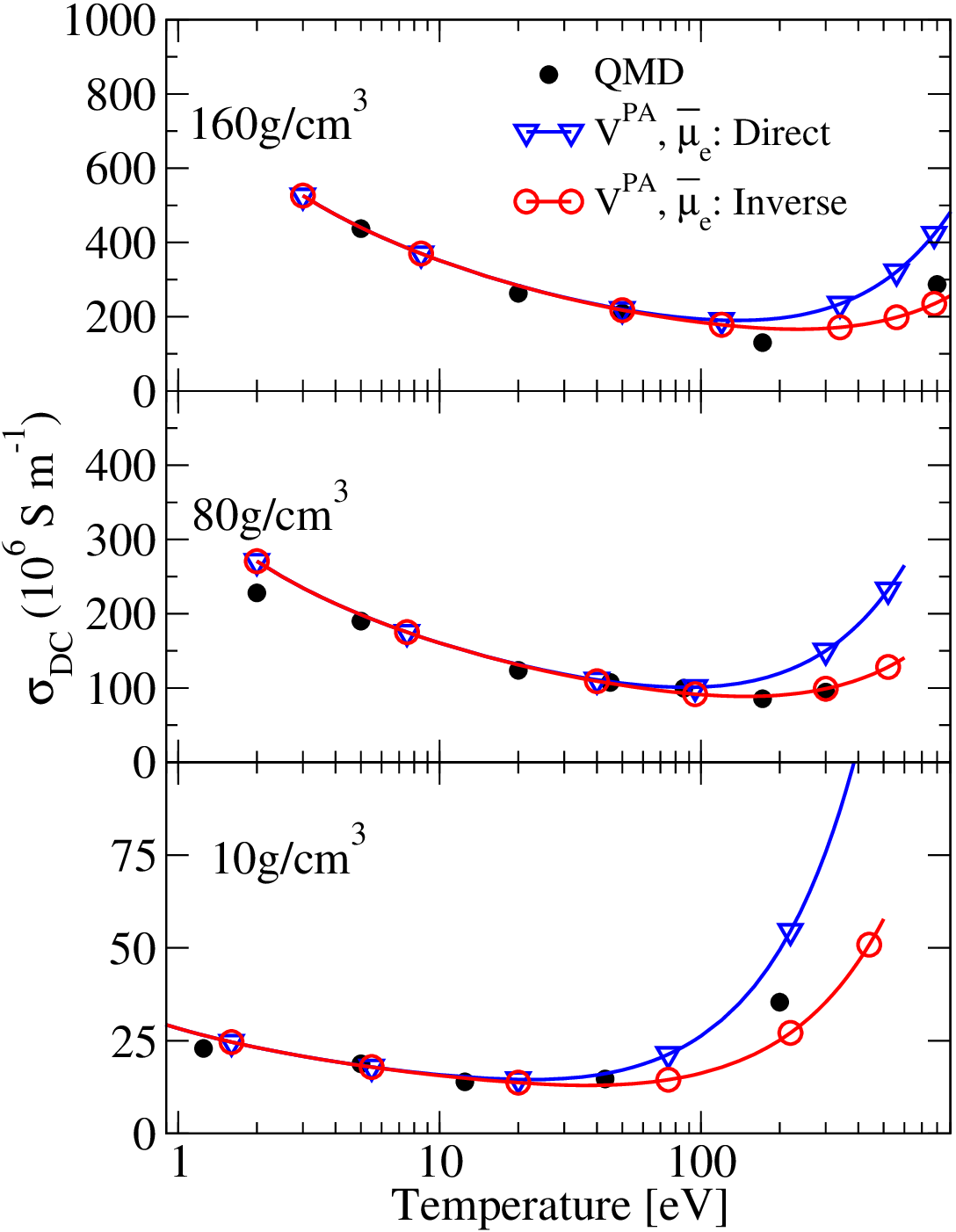}
\end{center}
\caption{(Color online) Electrical conductivity of warm dense hydrogen comparing
the direct approach (equation (\ref{direct}))
to the inverse approach (equation (\ref{inverse})).  All curves use PAMD-KS.
}
\label{fig_direct}
\end{figure}

\begin{figure}
\begin{center}
\includegraphics[scale=0.4]{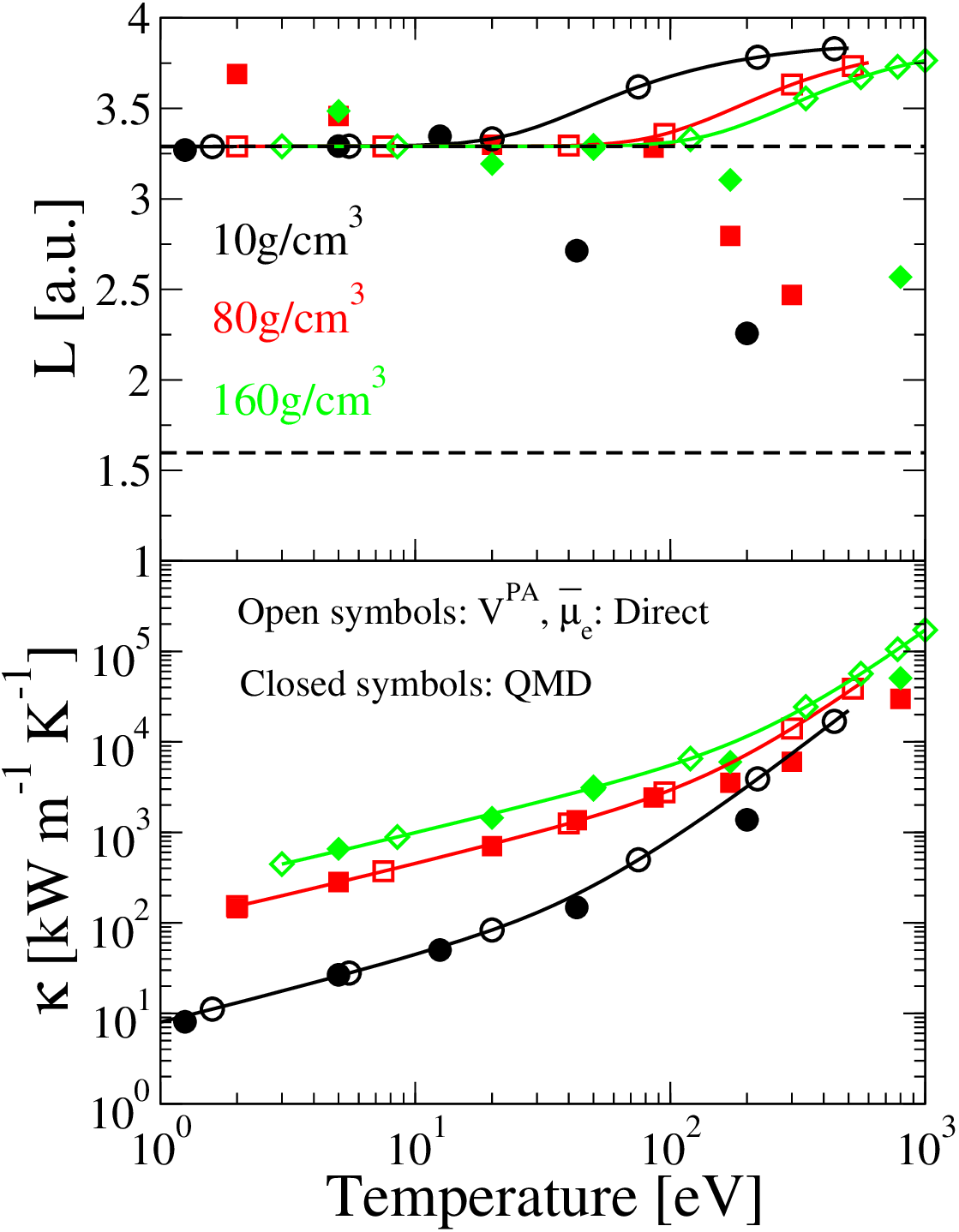}
\end{center}
\caption{(Color online) Thermal conductivity and Lorentz number of warm dense hydrogen using
the direct approach (equation (\ref{direct})).  All curves use PAMD-KS.  The horizontal dashed
lines indicate the degenerate ($L$ = 3.29 a.u.) and non-degenerate ($L$ = 1.597 a.u.) limits.
}
\label{fig_thermal}
\end{figure}

\subsection{Thomas-Fermi results\label{sec_tf}}
As noted earlier, PAMD is a density functional theory based model.  All
results presented up to this point have used the Kohn-Sham (KS) functional.
In figure \ref{fig_tf} we compare conductivities for warm dense aluminum
using KS to Thomas-Fermi (TF) calculations.  All results use the $V^{PA}$
potential and the $t$-matrix approach.  We show results from both chemical potential options.
Generally the TF model is remarkably close to the KS based result.  The results
are closer when using $n_e^0$ compared to $\bar{n}_e^0$.  This not surprising
since the lack of atomic shell structure in TF leads to a continuous change in the 
number of valence electrons with density and temperature.  Whereas, for KS 
the shell structure shows up as plateaus in the number of valence electrons.
On the other hand the TF estimation of
the chemical potential $\mu_e$ is generally quite close to the KS value, hence
the closer agreement.
Though TF is not as accurate a model as KS, it very computationally inexpensive, typically
taking less than 1 minute per density and temperature point, compared to roughly
30-60 minutes for Kohn-Sham calculations on the same modern, desktop computer.
It is therefore an attractive option for rapidly generating tables of 
reasonably accurate electrical conductivities.

\subsection{Direct approach and thermal conductivity \label{sec_direct}}
In this section we compare the direct and inverse approaches to 
conductivity (equations (\ref{direct}) and (\ref{inverse})).  We noted
earlier that as $T \to 0$ both approaches give the same result,
but at high $T$ the results will be different in general.  In
figure \ref{fig_direct} we compare these approaches to QMD data
for dense hydrogen.  As expected both methods give similar answers for low
temperature.  At higher $T$ the direct approach gives much large values for the
conductivity than the inverse approach.  In comparison
to the QMD data the inverse approach is generally closer, particularly
at 80 g/cm$^3$.  We have also checked the comparison to QMD for other
cases (beryllium and molybdenum) but the results from the direct and inverse methods
are similar due to the relatively low temperatures where QMD data is available.

In figure \ref{fig_thermal} we show the thermal conductivity and corresponding
Lorentz number
\begin{equation}
L = \frac{\kappa}{T \sigma_{DC}}
\end{equation}
calculated using the direct approach (see equation (\ref{kappa})).  At low
temperatures the agreement with QMD is excellent and the degenerate limit
of the Lorentz number is reproduced.  However, for higher temperatures
the agreement with QMD becomes poor, and the direct approach results in
an incorrect limit of the Lorentz number.  Such behavior was also found
in \cite{starrett12a}.  In that work the conductivities and Lorentz number
were calculated in a similar approach, the major differences being
that the structure factor was implicitly set to 1 and an average atom potential
and chemical potential were used.  We conclude that it is the direct approach,
as opposed to the scattering potential, that
is responsible for the incorrect limit of the Lorentz number.  Therefore
the direct approach must be treated with caution for higher temperatures where
the plasma is transitioning to being non-degenerate.

\section{conclusions\label{sec_conc}}
Expressions for electron transport coefficients have been systematically
tested by comparing to accurate but expensive Quantum Molecular Dynamics
(QMD) simulations.  A number of implementation options have been
tried, including choice of scattering potential, chemical potential and
ion-ion structure factor, all of which are generated by a self-consistent plasma model
called pseudo-atom molecular dynamics \cite{starrett15}.  
For the scattering potential, three types of potential were tested: a muffin-tin
like average atom potential; the potential due to one pseudo-atom (which is not
muffin-tin like); and a pseudopotential method.  We found that the pseudo-atom
potential gives consistently good agreement with the QMD data for a range of elements (including mixtures),
densities and temperatures.  The average atom potential gives good qualitative agreement 
but can be significantly in error for magnitude.  Finally, the pseudo-potential method
was found to be unreliable, and the reasons for this are discussed.  The effect of
two reasonable choices of chemical potential was tested:  first, a chemical potential
that recovers the expected number of valence electrons with a free electron density of
states ($\bar{\mu}_e$); and second, a chemical potential that recovers the expected number of valence electrons
with a calculated, non-free electron density of states ($\mu_e$).  The effect of the choice was generally
quite modest, but using $\bar{\mu}_e$ generally results in better agreement with the QMD
data.

Pseudo-atom molecular dynamics is a density functional theory based model and can therefore
use different approximations for the electronic kinetic energy functional.  We have compared
the effect of using the accurate Kohn-Sham approximation to the less accurate, but very computationally efficient,
Thomas-Fermi functional for warm dense aluminum.  In general we find the Thomas-Fermi results
are close to those using the Kohn-Sham functional, and its use therefore allows very rapid generation of 
electron conductivities that are reasonably accurate.

Finally, we have tested an approach in which the conductivity is calculated directly and compared 
to the inverse approach, where the resistivity is directly calculated.  Formally and numerically,
both give similar answers for degenerate plasmas, but for higher temperatures  where the plasma is
transitioning to the non-degenerate regime, they give significantly different
results.  For dense hydrogen the inverse approach was found to give conductivities that agree 
better with QMD data at high temperature.  We used the direct approach to calculate the thermal
conductivity and Lorenz number for warm dense hydrogen and found that, while good at low temperature
in the degenerate regime, at higher temperatures the method does not tend to the correct non-degenerate
limit, and significantly disagrees with QMD data.  As a result we conclude that the direct approach should be used
with caution in the non-degenerate regime.

\section*{Acknowledgments}
We thank S. B. Hansen for useful conversations and suggestions.
This work was performed under the auspices of the United States Department of Energy under contract DE-AC52-06NA25396
and LDRD number 20150656ECR.

\bibliographystyle{unsrt}
\bibliography{phys_bib}

\end{document}